\documentclass[conference]{IEEEtran}
\IEEEoverridecommandlockouts

\usepackage[T1]{fontenc}
\usepackage{cite}
\usepackage{amsmath,amssymb,amsfonts}
\usepackage{algorithmic}
\usepackage{graphicx}
\usepackage{textcomp}
\usepackage{xcolor}
\usepackage{booktabs}
\usepackage{url}
\usepackage{array}
\usepackage{enumitem}
\setlist[itemize]{leftmargin=*, labelsep=0.4em}
\usepackage{tikz}
\usetikzlibrary{positioning,fit,arrows.meta}
\newcommand{\metric}[1]{\multicolumn{1}{c}{#1}}
\def\BibTeX{{\rm B\kern-.05em{\sc i\kern-.025em b}\kern-.08em
    T\kern-.1667em\lower.7ex\hbox{E}\kern-.125emX}}
\begin{document}

\title{Build or Adopt? Custom-Built vs. General-Purpose LLM Agent Harnesses for Automotive GUI Testing}

\author{\IEEEauthorblockN{Arsham Gholamzadeh Khoee\IEEEauthorrefmark{1}\IEEEauthorrefmark{2},
Ruyu Fu\IEEEauthorrefmark{1},
Jiaqi Wei\IEEEauthorrefmark{1},
Ruikun Deng\IEEEauthorrefmark{2},\\
Robert Feldt\IEEEauthorrefmark{1},
Dhasarathy Parthasarathy\IEEEauthorrefmark{2},
Yinan Yu\IEEEauthorrefmark{1}\IEEEauthorrefmark{2}}
\IEEEauthorblockA{\IEEEauthorrefmark{1}\textit{Department of Computer Science and Engineering, Chalmers University of Technology}, Gothenburg, Sweden}
\IEEEauthorblockA{\IEEEauthorrefmark{2}\textit{Volvo Group}, Gothenburg, Sweden}
\IEEEauthorblockA{\{khoee, robert.feldt, yinan\}@chalmers.se, \{ruyu, jiaqiwe\}@student.chalmers.se,\\ \{ruikun.deng, dhasarathy.parthasarathy\}@volvo.com}
}

\maketitle

\begin{abstract}
As general-purpose LLM agent harnesses mature, organizations must decide whether to continue developing in-house systems or adopt existing alternatives. We study this decision at Volvo Group for GUI testing of Android-based in-vehicle infotainment (IVI) systems. We compare a production, custom-built harness with two general-purpose harnesses, Claude Code and OpenCode, keeping the tools, accessed through the Model Context Protocol (MCP), and the models the same. We evaluate the harnesses with up to four models on Volvo's IVI test cases and on open-ended tasks in public apps (Google Maps, Booking.com), and interview ten practitioners: developers, the product owner, and testers. On Volvo's stepwise IVI tests, general-purpose harnesses match the custom-built harness in success but vary more across runs. Their inference cost per task is 2.6--10.7 times higher, yet the engineering effort they save outweighs this: for a nightly IVI regression suite, total monthly cost, including engineering effort and inference, is 3--66\% lower. On open-ended public-app tasks, they succeed more often, mainly because they recover better from incorrect actions. Interviews suggest that external advances can turn an in-house harness into technical debt even without deterioration. Adopting a general-purpose harness shifts maintenance from the agent loop to tools and instructions, but gives teams less control over what the model receives. Regardless of the harness, testers want to inspect the agent's actions and ensure that it follows the prescribed test scenario.
\end{abstract}


\vspace{-3mm}
\section{Introduction}
\label{sec:intro}

Graphical user interface (GUI) testing of Android-based in-vehicle infotainment (IVI) systems is labor-intensive. These systems are stateful, depend on the state of the vehicle, and change with every software release; scripted tests break whenever the layout changes~\cite{coppola2017scripted}, and manual testing does not keep pace. Large language model (LLM) agents address this by executing natural-language test intents directly on the device, interpreting the screen rather than relying on static identifiers. When LLM agents first emerged, no capable general-purpose \emph{harness}, i.e., the software that turns a foundation model into an agent, existed, forcing organizations like Volvo to build their own, as for automotive software release analytics~\cite{khoee2026gatelens}. These in-house systems are \emph{custom-built}, relying on code to strictly define the agent's control loop, prompt construction, and subagent routing.

Research on LLM agents for mobile GUIs has followed the same route. Testing agents such as GPTDroid~\cite{liu2024make} and DroidAgent~\cite{yoon2024intent} explore apps or pursue test intents through purpose-built memory and planning, and task-automation agents such as AutoDroid~\cite{wen2024autodroid}, AppAgent~\cite{zhang2025appagent}, and Mobile-Agent~\cite{wang2024mobile} engineer UI representations, app-specific knowledge, and dedicated planning, decision, and reflection agents. These works evaluate new harness designs on benchmarks; to our knowledge, none compares a custom-built harness with a general-purpose one or examines what maintaining it entails in industrial practice.

Building such a harness is an investment in technical credit~\cite{bucaioni2026technical}, i.e., a design decision that requires higher initial effort in return for long-term benefits, but its value depends on an ecosystem that changes quickly. New model generations alter the agent's behavior, new requirements, such as operating several devices in one test, go beyond the original design, and both require revising the harness. Meanwhile, general-purpose harnesses, such as Claude Code~\cite{claudecode} and OpenCode~\cite{opencode}, have matured. They run a generic agent loop maintained by their vendors and can reach domain-specific tools through the Model Context Protocol (MCP)~\cite{mcpspec}, which would shift the in-house workload from maintaining an agent loop to maintaining tools and instructions. Thus, technical credit can turn into technical debt without any decay of the system itself. As a result, organizations with in-house agents face a build-versus-adopt decision. However, it is unknown whether a general-purpose harness can replace a custom-built one in industrial testing, what the move costs in operation, and what it changes for the engineers who maintain the agent and the testers who rely on it.

We study this decision at Volvo Group. Our key design choice is to vary only the harness: we keep the tools and device access of Volvo's production agent, use them in both the in-house custom-built harness and vendor-maintained general-purpose ones, such as Claude Code and OpenCode, which access them through an MCP server, and run all harnesses with the same models. As a result, differences between the setups reflect the harness rather than the tools or the model. We evaluate the setups across four models on 66 tasks derived from Volvo's IVI test cases and on 56 open-ended tasks from the AMEX dataset~\cite{chai2025amex}, repeat every run to measure flakiness, and interview ten practitioners: four developers, the product owner, and five testers. We address four research questions (RQs):

\begin{itemize}
    \item \textbf{RQ1 (Effectiveness):} How effectively do general-purpose harnesses execute intent-driven Android GUI tests compared with an in-house, custom-built harness using the same models?
    \item \textbf{RQ2 (Lock-in and robustness to change):} How robust is each harness's effectiveness, without adaptation, to a change of the underlying model or the application under test?
    \item \textbf{RQ3 (Maintenance and cost):} How does transitioning from an in-house, custom-built harness to a general-purpose one change the team's maintenance responsibilities and technical debt, and at what operational cost?
    \item \textbf{RQ4 (Stakeholder perspectives):} What do the stakeholders of the in-house testing agent value in it, and which of these qualities are preserved or put at risk by moving to a vendor-maintained harness?
\end{itemize}


This paper makes the following contributions:
\begin{itemize}
    \item An industrial comparison that varies only the harness: a production, custom-built harness against two general-purpose harnesses with the same tools and models, on 66 IVI and 56 public-app tasks, covering effectiveness, flakiness, robustness to model and application change, latency, and inference cost (RQ1–RQ3; Sections~\ref{sec:system_architecture} and~\ref{sec:quant_eval}).
    \item An interview study with ten practitioners (developers, the product owner, and testers) that identifies ecosystem-induced technical debt and shows how adopting a general-purpose harness redistributes maintenance and control (RQ3, RQ4; Section~\ref{sec:qual_eval}).
    \item A total-cost estimate combining engineering effort and inference cost, and lessons for practitioners who decide whether to keep or replace an in-house agent harness  (Section~\ref{sec:discussion}).
\end{itemize}

\section{Problem Formulation}
\label{sec:problem_formulation}

We model autonomous Android GUI testing as a sequential decision process~\cite{sutton1998reinforcement} in which an agent executes a natural-language test intent $G$ through a sequence of UI actions. At step $i$, the UI state $S^{(i)} = \langle V^{(i)}, X^{(i)} \rangle$ consists of a visual state $V^{(i)}$, the rendered screen captured as a screenshot, and a structural state $X^{(i)}$, the XML view hierarchy, whose nodes carry semantic properties such as \texttt{content-desc}, \texttt{resource-id}, and \texttt{class}.

The central difficulty is relating the two representations: tool calls target elements of $X^{(i)}$, whereas the agent interprets the screen through both. A visible element may carry no semantic text in the hierarchy, such as an icon in an empty \texttt{FrameLayout}; the hierarchy may contain nodes that are off-screen or covered; and the two representations can disagree about an element's state, e.g., whether a media control is playing. In such cases, neither representation alone suffices, and a mismatch leads the agent to act on the wrong element or misjudge the state of the app.

An action $a_i = \langle \tau, \rho \rangle \in \mathcal{A}$ combines an action type $\tau \in \{\text{TAP}, \text{SCROLL}, \text{TYPE}, \text{PRESS\_BACK}, \dots\}$ with its parameters $\rho$, e.g., coordinates derived from an element's bounding box in $X^{(i)}$. A policy $\pi$ selects the next action from the current state, the intent, and the history $H^{(i)} = [(S^{(0)}, a_0), \dots, (S^{(i-1)}, a_{i-1})]$, i.e., $a_i = \pi(S^{(i)}, G, H^{(i)})$. Because Android UIs are asynchronous, the transition $\mathcal{T}(S^{(i+1)} \mid S^{(i)}, a_i)$ is stochastic: if the agent acts before an animation or screen transition has finished, it may reach an unintended state. This stochasticity is one source of the run-to-run variation that we measure as flakiness (Section~\ref{sec:quantitative}).

\section{System Architecture}
\label{sec:system_architecture}

\begin{figure*}[t]
\centering
\includegraphics[width=\textwidth]{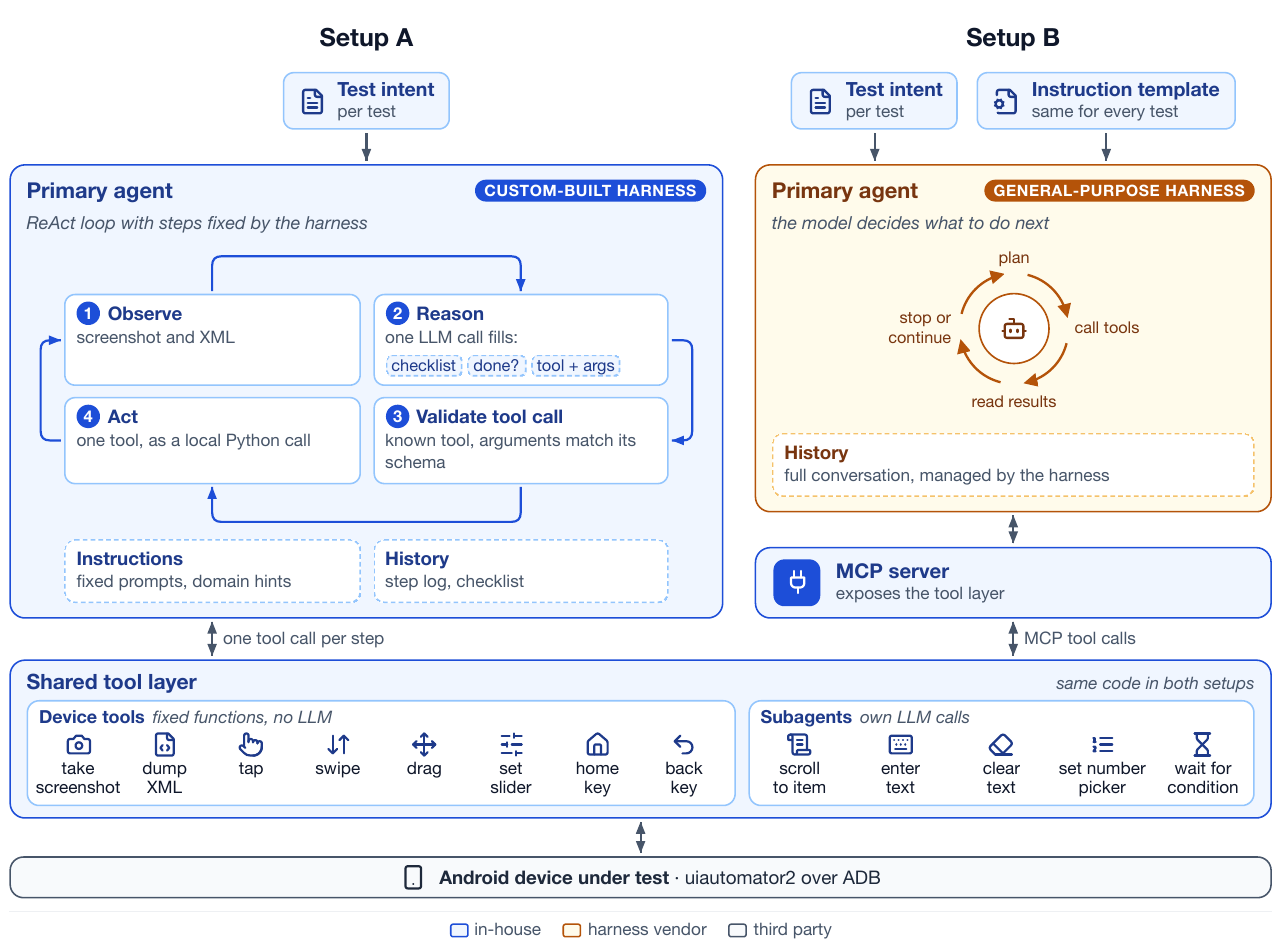}
\caption{Architecture of the two setups. In Setup~A, the primary agent executes each test intent in a ReAct loop whose steps (1--4) are fixed by the custom-built harness. In Setup~B, the primary agent, built on a general-purpose harness, decides by itself how to proceed within a generic loop and is adapted to the task through outer-harness components. Both setups use the same tool layer and device access. Colors indicate who maintains each component.}
\vspace{-3mm}
\label{fig:architecture}
\end{figure*}

In this section, we describe the testing system that Volvo has developed and used in production, and the direction we explore for its evolution. We use the term \emph{harness} for the software around a foundation model that turns it into an agent~\cite{meng2026agent}: the control loop, the prompts, the context given to the model, and the tools it can use. In a \emph{custom-built harness}, the harness code defines the workflow: the steps, the routing between agents, and which specialist agent runs when; the model works inside decision slots designed by the developers. In a \emph{general-purpose harness}, a single generic loop with broad tools lets the model decide what to do next and when to delegate work to subagents; such a harness is adapted to a domain not by changing its loop, but through outer-harness components such as instruction files, MCP servers, and skills. The two kinds differ in where the decisions are made: in the harness code or by the model. Setup~A, which we designed and deployed for intent-driven GUI testing at Volvo, is based on a custom-built harness (Section~\ref{sec:arch_inhouse}). Setup~B keeps the tools of Setup~A but replaces its harness with a general-purpose one, that of Claude Code~\cite{claudecode} or OpenCode~\cite{opencode}, configured through an instruction template and an MCP~\cite{mcpspec} server (Section~\ref{sec:arch_mcp}). Fig.~\ref{fig:architecture} shows both setups. As the tools are the same, both setups share the state representation $S^{(i)}$ and the action space $\mathcal{A}$ of Section~\ref{sec:problem_formulation}; they differ in the policy $\pi$ that selects the next action and in who maintains the code that implements it (Section~\ref{sec:arch_ownership}).

\subsection{Setup A: Based on a Custom-Built Harness}
\label{sec:arch_inhouse}

Setup~A is in production use at Volvo: testers write their test cases against it, and an internal team of developers maintains it. Its primary agent follows the ReAct pattern~\cite{yao2022react} in a loop of four fixed steps (Fig.~\ref{fig:architecture}, left). In each iteration, it takes exactly one action, returns its decision in a typed format that includes an explicit checklist, and reports completion through a dedicated output field. This prescribed loop is what distinguishes the custom-built harness from a general-purpose one.

\emph{Step~1: Observe.} The harness captures a screenshot and dumps the view hierarchy through device tools (Section~\ref{sec:arch_tools}). It simplifies the hierarchy by keeping only nodes that are interactive or display text, and only the attributes needed to identify an element and judge its state: text, content description, resource identifier, bounds, and the flags \emph{checked}, \emph{selected}, and \emph{enabled}. The primary agent is instructed to judge an element's state from these flags rather than from the screenshot, and to verify it with a tool when the two disagree. The harness does not align visual and structural elements explicitly; it provides both to a multimodal model and relies on the model's reasoning to align them.

\emph{Step~2: Reason.} A single LLM call returns an updated checklist that decomposes the intent into atomic UI steps~\cite{khoee2024gonogo}, a judgment of whether the intent is fulfilled, and the next tool with its arguments. All LLM calls in the harness use zero-shot chain-of-thought prompting~\cite{wei2022chain,kojima2022large}.

\emph{Steps~3--4: Validate and act.} The harness checks that the selected tool exists and validates its arguments against a schema derived from the tool's signature. An invalid call is recorded as a failed step; a valid one is invoked as a local Python function of the shared tool layer, and its outcome is recorded in the step log.

\emph{Instructions and history.} The harness rebuilds the prompt from scratch at every iteration. Besides the intent and the current observation, i.e., the screenshots before and after the last action, the simplified view hierarchy, and the foreground app and screen, the prompt contains instructions and history. The \emph{instructions} are fixed: the system prompts, the tool descriptions, and a list of domain hints, such as how long to wait for a pop-up dialog. The \emph{history} consists of the step log, which records the purpose, tool, arguments, and outcome of every previous action and realizes $H^{(i)}$, and the checklist. Thus, the harness itself decides what the model sees: earlier steps survive only as entries in the step log, and only the screenshots around the last action are included.

\emph{Run records.} For every run, the harness produces a report with the screenshot, reasoning, checklist, and action of each step, a log of every prompt and response, and a trajectory that links the primary agent's run to those of its subagents. These records are the main means of diagnosing whether a failed run reveals an app defect or an agent mistake.

\subsection{Setup B: Based on a General-Purpose Harness}
\label{sec:arch_mcp}

We explore Setup~B to reduce Setup~A's maintenance burden. We keep the tool layer and device access of Setup~A and replace only its primary agent with one built on the harness of Claude Code or OpenCode. Both were built for software development, but their harness is not specific to coding. Setup~B adapts it through two outer-harness components: an instruction template and an MCP server through which the primary agent reaches the shared tool layer (Fig.~\ref{fig:architecture}). The instruction template is identical for every test and shared by both harnesses. Besides a few general rules about our test environment, such as which device to use and that tasks run in installed apps rather than in a browser, it prescribes a general workflow: observe the screen before acting, take element bounds from the hierarchy rather than guessing coordinates, and perform and verify one action at a time.

\emph{Primary agent.} The primary agent receives the intent, the tools, and the instructions, and decides by itself how to accomplish the task. Its harness imposes no fixed steps or decision format: at each turn, the model calls the tools it chooses, reads their results, and decides whether to continue, including when to observe the screen again and when the task is complete. The history of the run is the conversation context, which accumulates every tool call and its result and which the harness condenses when it grows long. Both the loop and the context management are maintained by the vendor and adapted to new models as they are released. Like Setup~A, Setup~B is a multi-agent system: the primary agent can delegate work to subagents of its own, and its intent-based tools are executed by the subagents of the shared tool layer.

\emph{MCP server.} An MCP server exposes tools to an agent, each with a name, a natural-language description, and a schema for its arguments. Our server exposes the shared tool layer (Section~\ref{sec:arch_tools}). Every action tool returns the simplified hierarchy before and after the action, so the agent can verify its effect without a separate observation call. The server reuses the tools of Setup~A but none of its orchestration: which tool to call, when, and when to stop are left to the primary agent.

\subsection{Shared Tool Layer and Device Access}
\label{sec:arch_tools}

The tool layer is the same code in both setups: Setup~A invokes its tools as local Python functions, and Setup~B through the MCP server. Both access the device through uiautomator2~\cite{uiautomator2}, which drives Android's UI Automator framework over the Android Debug Bridge (ADB).

\emph{Device tools} interact with the device without involving an LLM. Two serve observation: one captures a screenshot, and one dumps the view hierarchy as an XML tree, which is simplified as in Setup~A; together, they provide $V^{(i)}$ and $X^{(i)}$. The others act through a single gesture or key event: tap, swipe, drag, set a slider, and press the home or back key. Tools that target an element take its bounding box from the hierarchy and compute the gesture coordinates themselves, e.g., the center of the box for a tap.

\emph{Subagents} perform subtasks that require several interactions or task-specific reasoning: scrolling to an item in a list, entering or clearing text, setting a number picker, and waiting until a condition holds on screen. The primary agent delegates such a subtask with a natural-language intent, such as ``scroll to find Developer options and tap it''. Each subagent has its own specialized prompt and LLM context and returns only the final result, which keeps intermediate screens out of the primary agent's context. Because subagents call an LLM, their behavior, unlike that of device tools, depends on the underlying model.

\subsection{Maintenance and Ownership}
\label{sec:arch_ownership}

Moving from Setup~A to Setup~B changes who maintains what. In Setup~A, Volvo maintains the entire stack, from the primary agent to the device layer. Its control loop, prompts, and construction of the context are designed and tuned together for the models in use. When the backbone model is replaced, the same harness can behave differently~\cite{chen2023chatgpt}: a new model may decompose an intent differently, declare completion earlier or later, or repeat an action where its predecessor did not. Therefore, keeping the harness effective across model updates requires revising these components, and this effort recurs with every model generation the team adopts. The same holds for new requirements: supporting a new feature or test concept requires the team to develop and extend the harness itself, i.e., its prompts, domain hints, and often its tool set or the routing between its agents.

In Setup~B, the agent loop, context management, and tool-calling protocol are maintained by the vendors of the general-purpose harness, while Volvo maintains the MCP server, the instruction template, and the tool layer, including the subagents behind the intent-based tools. When a new model is adopted, the vendor's harness absorbs most of the adaptation; what remains in-house are the instruction template, the tool descriptions, and the subagents. Likewise, a new requirement is usually addressed by updating the instruction template, and the primary agent adapts its behavior accordingly. Only a new device capability requires adding a tool to the shared tool layer, which the MCP server then exposes without any change to the harness. Because the tool layer is identical, differences between the two setups reflect the policy $\pi$ and the orchestration, not the observation or actuation code.

\section{Quantitative Evaluation}

\label{sec:quant_eval}

This section compares how well the two setups execute test intents and what their execution costs. After describing the setup (Section~\ref{sec:quantitative}), we report effectiveness, i.e., how often and how reliably each configuration achieves the intent, which answers RQ1 (Section~\ref{sec:rq1}), robustness to a change of model or application, which answers RQ2 (Section~\ref{sec:rq_robust}), and efficiency, i.e., the latency, tokens, and inference cost per task, which answers the operational side of RQ3 (Section~\ref{sec:rq3_quant}).

\subsection{Setup}

\label{sec:quantitative}

\subsubsection{Configurations}

Setup~A is run with GPT-5.1, the model it uses in production, as the production baseline. To separate the effect of the harness from that of the model, both setups are also run with the same models: GPT-5.4, Claude Sonnet~4.6, and Claude Opus~4.6, the latter on Google Maps only. Claude Code is used only with Anthropic models.

\subsubsection{Benchmarks}

We report on two benchmarks. The \emph{Volvo IVI benchmark} consists of 66 tasks on the Infotainment Control Module (ICM) of a Volvo test rig, covering climate control, lights, driving assistance, and system and vehicle settings. Its intents are written as stepwise tests, e.g., ``Open the Climate control app; turn the floor air distribution on; turn it off'', and are derived from Volvo's existing test cases. The \emph{AMEX benchmark} consists of 56 tasks from the open-source AMEX dataset~\cite{chai2025amex}, 40 in Google Maps and 16 in Booking.com, executed on Android phones; its intents are single goals, such as searching for a hotel with given dates and filters. We selected these two applications because their tasks could be adapted to dynamic settings similar to our industrial context and complement each other: map search and navigation in Google Maps, and long, multi-step form filling in Booking.com. To evaluate how a run reaches its goal, we also use the milestones of each task, i.e., natural-language descriptions of intermediate UI states.

\subsubsection{Execution and Evaluation}

All configurations use the same tasks, devices, and tool layer. Before each task, we return the device to its home screen and launch the target app. Each configuration is run three times on AMEX and twice on the Volvo rig. Every run is recorded as a trace of tool calls, outputs, and screenshots, which is scored by an LLM judge~\cite{zheng2023judging,li2025generation,gu2026survey}, GPT-5.6-sol. To check its verdicts across model families, we additionally scored the Booking.com runs with Claude Opus~4.8; the two judges agree on 93.4\% of the runs (Cohen's $\kappa = 0.85$~\cite{cohen1960coefficient}) and rank the setups identically. To validate the judge against human judgment, we manually labeled task success for one repetition of the Booking.com benchmark. On the 88\% of runs with an unambiguous outcome, GPT-5.6-sol agreed with our manual labels in every case across both setups. The remaining 12\% belong to three tasks with debatable success criteria (stopping at a final payment or personal-details screen, or applying valid filters that yielded no matching offers). These ambiguous runs account for all but one of the disagreements between the two judges in this repetition (in the remaining case, our manual check confirmed GPT-5.6-sol), and because GPT-5.6-sol counted all but one of them as failures in both setups, they do not affect the comparison. Runs that end in a timeout or a runtime error count as failures. To check whether differences between setups using the same model are statistically significant, we compare their task success and cost per task using an exact paired Wilcoxon signed-rank test~\cite{wilcoxon1945individual}.

\subsubsection{Metrics}
We measure effectiveness (RQ1, RQ2) and efficiency (RQ3) per run:

\begin{itemize}
    \item \emph{Task success:} whether the run achieved the intent, judged from the full trace, the last three screenshots, and the agent's reasoning.
    \item \emph{Milestone reach:} the fraction of milestones reached, judged from the recorded view hierarchies and screenshots.
    \item \emph{Tool correctness:} the fraction of action tool calls judged correct, given the view hierarchy before and after each call.
    \item \emph{Fault recovery:} among runs with at least one tool call not judged correct, the fraction that still succeeded.
    \item \emph{Flakiness:} the fraction of tasks whose success outcome differs across the repeated runs of a configuration~\cite{luo2014empirical}. As the number of runs differs across benchmarks, we compare flakiness only within each benchmark.
    \item \emph{Efficiency:} latency, number of tool calls, tokens, and total inference cost in USD per task, including the LLM calls of the subagents.
\end{itemize}

\begin{figure*}[t]
\centering
\includegraphics[width=\textwidth]{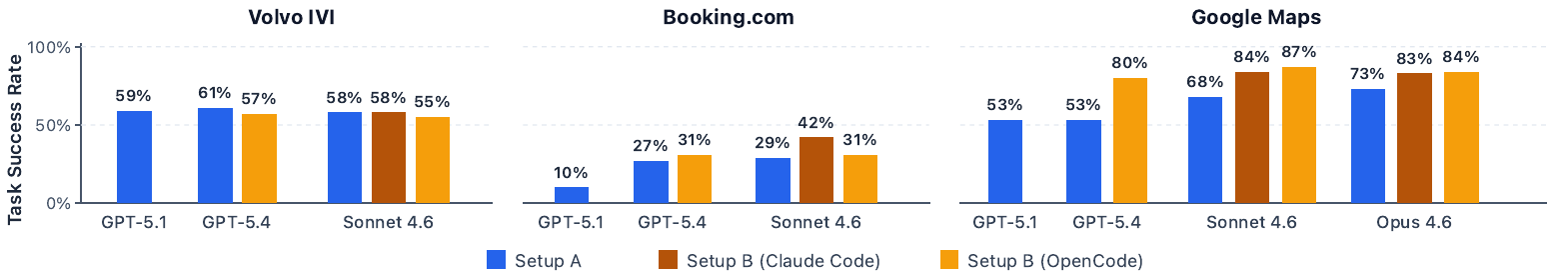}
\caption{Task success per benchmark and configuration, grouped by model.}
\label{fig:success}
\end{figure*}

\begin{table*}[t]
\caption{Effectiveness per benchmark and configuration beyond task success (Fig.~\ref{fig:success}). Bold marks the best value per model and benchmark; lower flakiness is better.}
\label{tab:effectiveness}
\centering
\footnotesize
\setlength{\tabcolsep}{3pt}
\resizebox{\textwidth}{!}{%
\begin{tabular*}{1.2\textwidth}{@{\extracolsep{\fill}}ll*{12}{c}@{}}
\toprule
 & & \multicolumn{4}{c}{Volvo IVI} & \multicolumn{4}{c}{Booking.com} & \multicolumn{4}{c}{Google Maps} \\
\cmidrule(lr){3-6}\cmidrule(lr){7-10}\cmidrule(l){11-14}
 & & \metric{Milestone} & \metric{Tool} & \metric{Fault} & & \metric{Milestone} & \metric{Tool} & \metric{Fault} & & \metric{Milestone} & \metric{Tool} & \metric{Fault} & \\
Model & Setup & \metric{reach} & \metric{correctness} & \metric{recovery} & \metric{Flakiness} & \metric{reach} & \metric{correctness} & \metric{recovery} & \metric{Flakiness} & \metric{reach} & \metric{correctness} & \metric{recovery} & \metric{Flakiness} \\
\midrule
GPT-5.1 & A & 0.87 & 0.97 & 0.46 & 0.03 & 0.39 & 0.89 & 0.10 & 0.06 & 0.72 & 0.97 & 0.40 & 0.40 \\
\midrule
GPT-5.4 & A & 0.87 & \textbf{0.96} & \textbf{0.25} & \textbf{0.02} & 0.59 & \textbf{0.92} & 0.18 & \textbf{0.19} & 0.74 & \textbf{0.93} & 0.24 & 0.45 \\
 & B (OpenCode) & \textbf{0.89} & \textbf{0.96} & 0.00 & 0.05 & \textbf{0.73} & \textbf{0.92} & \textbf{0.28} & 0.31 & \textbf{0.84} & 0.89 & \textbf{0.78} & \textbf{0.23} \\
\midrule
Sonnet 4.6 & A & \textbf{0.90} & 0.93 & \textbf{0.24} & \textbf{0.02} & 0.56 & 0.91 & 0.12 & 0.38 & 0.82 & \textbf{0.94} & 0.40 & 0.35 \\
 & B (Claude Code) & \textbf{0.90} & \textbf{0.96} & 0.10 & 0.09 & \textbf{0.78} & \textbf{0.93} & \textbf{0.38} & 0.50 & 0.88 & 0.88 & \textbf{0.83} & 0.25 \\
 & B (OpenCode) & 0.89 & 0.94 & 0.15 & 0.05 & 0.72 & \textbf{0.93} & 0.27 & \textbf{0.19} & \textbf{0.90} & 0.89 & \textbf{0.83} & \textbf{0.13} \\
\midrule
Opus 4.6 & A & -- & -- & -- & -- & -- & -- & -- & -- & 0.84 & \textbf{0.98} & 0.53 & 0.48 \\
 & B (Claude Code) & -- & -- & -- & -- & -- & -- & -- & -- & \textbf{0.87} & 0.89 & 0.78 & \textbf{0.25} \\
 & B (OpenCode) & -- & -- & -- & -- & -- & -- & -- & -- & 0.86 & 0.89 & \textbf{0.81} & \textbf{0.25} \\
\bottomrule
\end{tabular*}}
\end{table*}

\begin{table*}[t]
\caption{Efficiency per task: mean latency (s), tool calls, tokens (thousands), and inference cost (USD). Token counts are comparable only within a model. Bold marks the lowest value per model and benchmark.}
\label{tab:efficiency}
\centering
\footnotesize
\setlength{\tabcolsep}{3pt}
\resizebox{\textwidth}{!}{%
\begin{tabular*}{1.2\textwidth}{@{\extracolsep{\fill}}ll*{12}{c}@{}}
\toprule
 & & \multicolumn{4}{c}{Volvo IVI} & \multicolumn{4}{c}{Booking.com} & \multicolumn{4}{c}{Google Maps} \\
\cmidrule(lr){3-6}\cmidrule(lr){7-10}\cmidrule(l){11-14}
Model & Setup & \metric{Latency} & \metric{Calls$^{a}$} & \metric{Tokens} & \metric{Cost} & \metric{Latency} & \metric{Calls$^{a}$} & \metric{Tokens} & \metric{Cost} & \metric{Latency} & \metric{Calls$^{a}$} & \metric{Tokens} & \metric{Cost} \\
\midrule
GPT-5.1 & A & 297 & 6.8 & 95 & 0.13 & 538 & 41.8 & 989 & 1.41 & 198 & 14.6 & 283 & 0.40 \\
\midrule
GPT-5.4 & A & 341 & \textbf{8.9} & \textbf{138} & \textbf{0.35} & 520 & 41.6 & \textbf{1\,014} & \textbf{2.67} & 245 & 22.6 & \textbf{465} & \textbf{1.23} \\
 & B (OpenCode) & \textbf{305} & 14.1 & 354 & 0.92 & \textbf{508} & \textbf{34.6} & 3\,625 & 9.06 & \textbf{207} & \textbf{12.8} & 937 & 2.34 \\
\midrule
Sonnet 4.6 & A & 390 & \textbf{8.6} & \textbf{132} & \textbf{0.46} & 647 & \textbf{32.8} & \textbf{967} & \textbf{3.53} & 311 & 17.0 & \textbf{461} & \textbf{1.57} \\
 & B (Claude Code) & \textbf{363} & 27.8 & 1\,639 & 4.97 & \textbf{603} & 48.7 & 4\,514 & 13.49 & 307 & 17.5 & 1\,701 & 5.01 \\
 & B (OpenCode) & 388 & 17.4 & 698 & 2.15 & 641 & 40.2 & 3\,477 & 10.40 & \textbf{256} & \textbf{14.5} & 1\,235 & 3.69 \\
\midrule
Opus 4.6 & A & -- & -- & -- & -- & -- & -- & -- & -- & 295 & 16.4 & \textbf{425} & \textbf{2.40} \\
 & B (Claude Code) & -- & -- & -- & -- & -- & -- & -- & -- & \textbf{237} & \textbf{14.3} & 1\,384 & 6.86 \\
 & B (OpenCode) & -- & -- & -- & -- & -- & -- & -- & -- & 286 & 14.8 & 1\,362 & 6.81 \\
\bottomrule
\end{tabular*}}

\smallskip
\raggedright\footnotesize $^{a}$For Setup~A, the screenshot and view-hierarchy dump that the harness performs at every step are not counted, since the model does not decide to make them.
\end{table*}

\subsection{Effectiveness (RQ1)}
\label{sec:rq1}

Fig.~\ref{fig:success} reports task success, Table~\ref{tab:effectiveness} reports the other effectiveness metrics, and Table~\ref{tab:efficiency} reports the efficiency metrics, each per benchmark and configuration. Configurations are grouped by model, so that each setup is compared with the others on the same model.

\emph{Volvo IVI benchmark: comparable success.} All configurations succeed in 55--61\% of the runs. With the same model, Setup~B differs from Setup~A by $-4.5$ to $-0.8$ points, and no difference is significant ($p \geq 0.17$). Milestone reach (87--90\%) and tool correctness (93--97\%) are equally close. Setup~B yields inconsistent outcomes on slightly more tasks (3--6 of 66, against 1--2 for Setup~A). The production configuration, Setup~A with GPT-5.1, performs comparably to newer models. 

\emph{AMEX benchmark: higher success of Setup~B.} On Google Maps, Setup~B outperforms Setup~A with every model: by 28~points with GPT-5.4, 17--19 with Sonnet~4.6, and 9--11 with Opus~4.6 ($p \leq 0.01$ for GPT-5.4 and Sonnet~4.6; $p = 0.04$ and $0.06$ for Opus~4.6). The gap is largest where Setup~A is weakest: with GPT-5.4, Setup~A succeeds in 53\% of the runs, with Opus~4.6 in 73\%. This suggests that the general-purpose harness compensates most where the model performs poorly within the fixed loop. Setup~B is also more stable there, with 13--25\% flaky tasks against 35--48\%, partly because its higher success rate leaves fewer tasks with mixed outcomes. Booking.com is the hardest benchmark, with success rates of 10--42\%. Setup~B is ahead with the same model, but the differences are not significant over 16 tasks. Nonetheless, its milestone reach is clearly higher (72--78\% against 56--59\%): Setup~B reaches more of the intermediate UI states on the way to the goal.

\emph{Recovery rather than precision.} Setup~B does not make fewer incorrect tool calls; on Google Maps, its tool correctness is even lower (88--89\% against 93--98\%). The difference lies in what follows an incorrect call: Setup~B still succeeds in 78--83\% of such runs on Google Maps and 27--38\% on Booking.com, while Setup~A succeeds in 24--53\% and 10--18\%. We attribute this to the accumulated context of Setup~B, which lets the agent notice when an action has no effect or the wrong effect and correct it, whereas Setup~A rebuilds its prompt at every step from a compact record of previous actions. On the Volvo IVI benchmark, this pattern does not hold: incorrect calls are rare (tool correctness 93--97\%), and fault recovery is low in all configurations. Because the IVI intents prescribe each step, an incorrect action takes the agent off the specified procedure with little room to recover, whereas the open-ended AMEX goals can often still be reached along another path.

The contrast between the benchmarks is plausibly explained by the form of the intents. The Volvo IVI intents are written as stepwise tests, so planning is largely done by the tester, and the prescribed loop of Setup~A suffices. The AMEX intents state only a goal, which requires the agent to plan, explore, and recover and thereby favors the autonomy of Setup~B.

\vspace{2mm}
\noindent\fbox{%
    \parbox{\dimexpr\linewidth-2\fboxsep-2\fboxrule}{%
        \textbf{RQ1 findings:}
        General-purpose harnesses provide the clearest benefit on tasks requiring planning and recovery. They achieve comparable success to the custom-built harness on prescribed IVI procedures, with somewhat less consistent outcomes across runs, but perform better on open-ended tasks. Their advantage lies in recovering from mistakes rather than consistently avoiding them.
    }%
}

\subsection{Robustness to Change (RQ2)}
\label{sec:rq_robust}

We examine how each harness's effectiveness holds up, without adaptation, under the two kinds of change that drive maintenance (Section~\ref{sec:arch_ownership}): a new model and a new application under test. Setup~A was refined for GPT-5.1 and Volvo's IVI applications and runs all other models and the AMEX applications unchanged, while Setup~B uses the same instruction template throughout. Table~\ref{tab:robustness} summarizes both dimensions.

\begin{table}[t]
\caption{Task success without adaptation. Model change: range across the models run with both setups (spread in points). Application change: Setup~B minus Setup~A with the same model (points).}
\label{tab:robustness}
\centering
\footnotesize
\setlength{\tabcolsep}{4pt}
\begin{tabular*}{\columnwidth}{@{\extracolsep{\fill}}lccc@{}}
\toprule
 & \multicolumn{2}{c}{Model change} & \metric{Application change} \\
\cmidrule(lr){2-3}\cmidrule(l){4-4}
Benchmark & \metric{Setup~A} & \metric{Setup~B$^{a}$} & \metric{B $-$ A} \\
\midrule
Volvo IVI & 58--61\% (3) & 55--57\% (2) & $-4.5$ to $-0.8$ \\
Booking.com & 27--29\% (2) & 31\% (0) & $+2.1$ to $+12.5$ \\
Google Maps & 53--73\% (21) & 80--87\% (7) & $+9.2$ to $+27.5$ \\
\bottomrule
\end{tabular*}

\smallskip
\raggedright\footnotesize $^{a}$OpenCode only; Claude Code runs only Anthropic models.
\end{table}

\emph{Model change.} On the Volvo IVI benchmark and Booking.com, neither setup depends much on the model: with the models used by both setups, success varies by at most 3 points. On Google Maps, success varies by 21 points in Setup~A (53--73\%) and by 7 points in OpenCode (80--87\%). This difference stems mainly from GPT-5.4, with which Setup~A succeeds in 53\% of the runs against 80\% for OpenCode; with Sonnet~4.6 and Opus~4.6, all three harnesses vary by at most 5 points.

\emph{Application change.} Setup~A performs as well as Setup~B only on the applications it was refined for: on both AMEX applications, it falls behind with every model, most clearly on Google Maps (Table~\ref{tab:robustness}). Its production model shows the same pattern: GPT-5.1, for which Setup~A was refined, performs comparably to newer models on the Volvo IVI benchmark but has the lowest success on Booking.com (10\%). As each comparison uses the same tasks and model, this shift reflects how well each harness holds up outside the domain it was refined for. However, the change of application coincides with a change from stepwise intents to open-ended goals and from the test rig to phones, so the effect cannot be attributed to the application alone.

\vspace{2mm}
\noindent\fbox{%
    \parbox{\dimexpr\linewidth-2\fboxsep-2\fboxrule}{%
        \textbf{RQ2 findings:}
         Both harness types tolerate most model changes, but the custom-built harness transfers less successfully beyond its original setting. Its comparable performance on IVI tests does not extend to the AMEX applications, suggesting that specialization limits reuse across applications and task styles.
    }%
}

\subsection{Efficiency (RQ3)}
\label{sec:rq3_quant}
\emph{Cost.} With the same model, Setup~B costs 1.9 to 10.7 times as much per task as Setup~A ($p < 0.001$ for every comparison). The gap is the largest on the Volvo IVI benchmark, where Claude Code with Sonnet~4.6 costs \$4.97 per task against \$0.46 for Setup~A. Token usage grows by similar factors (2.0 to 12.5). In Setup~A, context stays concise because the harness rebuilds the prompt at every step from a compact record of previous actions and the current screen. In Setup~B, the conversation context grows with every tool call and its output, including the view hierarchies returned before and after each action, and is sent to the model again at every turn; most of its tokens are therefore re-sent as context. With the same model, OpenCode costs less than or about as much as Claude Code, with similar success on the Volvo IVI benchmark and Google Maps. The production configuration is the cheapest on every benchmark, at \$0.13 per task on the Volvo IVI benchmark.

\emph{Latency.} Despite its larger contexts, Setup~B is not slower: with the same model, its mean latency per task is 3 to 57~s lower than that of Setup~A. On the Volvo IVI benchmark, Setup~A issues fewer tool calls than Setup~B but still takes longer, because each of its steps takes longer. Every step of Setup~A begins with capturing a screenshot and dumping the view hierarchy. This observation is a fixed part of the step, performed deterministically by the harness; it is not counted as a tool call, since the model does not decide to make it. Setup~B, in contrast, observes the screen only when the agent decides to, and these observations are counted as tool calls.

\vspace{2mm}
\noindent\fbox{%
   \parbox{\dimexpr\linewidth-2\fboxsep-2\fboxrule}{%
       \textbf{RQ3 findings (cost, quantitative):}
        Moving to a general-purpose harness does not increase latency, but it raises the inference cost per task by a factor of 2.6 to 10.7 on the IVI benchmark (1.9 to 10.7 across all benchmarks) with the same model. However, for a nightly IVI regression suite, the saved engineering effort (0.5 $\rightarrow$ 0.1 full-time equivalents,  Section~\ref{sec:discussion}) outweighs this: total monthly cost, including engineering effort and inference, is 3--66\% lower.
   }%
}

\section{Qualitative Evaluation}
\label{sec:qual_eval}

This section examines the maintenance side of RQ3 and the stakeholder perspectives of RQ4. We first describe how we conducted and analyzed the interviews (Section~\ref{sec:qualitative}) and then report our findings (Section~\ref{sec:qual_findings}).

\subsection{Method}
\label{sec:qualitative}

\subsubsection{Participants}
We interviewed ten practitioners at Volvo from the three stakeholder groups of Setup~A: four developers who build and maintain it (D1--D4), its product owner (PO), and five quality assurance (QA) testers who are actively using it for conducting tests (T1--T5). Three developers had direct experience with both setups, while D1 primarily worked on Setup~A and had limited exposure to Setup~B. The PO was involved in planning and coordinating the evolution toward Setup~B. Since Setup~B had not yet been released to testers, their interviews focused on the operational properties they value in Setup~A and the requirements a replacement should preserve.

\subsubsection{Data Collection}
We conducted semi-structured interviews~\cite{runeson2009guidelines}, each lasting 50--60 minutes. To situate both setups, we first walked participants through the evolution of GUI testing at the company, from manual and scripted testing, including AI-assisted scripting, to an agent built on an in-house, custom-built harness (Setup~A) and one built on a general-purpose harness (Setup~B). We then presented eleven properties tailored to the participant's role, whose ratings served to structure the discussion rather than as measurements. Developers rated engineering properties, i.e., build effort, maintainability, debuggability, lock-in, integration friction, extensibility, reproducibility, cost, latency, evaluability, and onboarding time, on a scale from 1 (least favorable) to 10 (most favorable). They followed an \emph{anchor-and-compare} strategy: they first rated each property for Setup~A, the anchor, and then for Setup~B. Testers rated user-facing properties of Setup~A on the same scale, i.e., execution time, coverage, consistency, reproducibility, false positives, false negatives, scenario coherence, verification effort, visibility, control, and onboarding time. The PO did not rate the properties but instead discussed which of the engineering properties matter most for product decisions and what they imply for the choice between the two setups. In the remainder of each interview, we discussed the ratings with the participant, asked for the reasons behind them, and, for developers, for the differences between the two setups. In addition, the team's manager estimated the share of developer time spent on maintaining and extending Setup~A.

\subsubsection{Data Analysis}
We analyzed the interviews through a lightweight thematic analysis~\cite{braun2006using,lenberg2024qualitative}. For each interview, we summarized the explanations provided for the ratings or, for the PO, for the properties discussed, and coded notable experiences and concerns. We then compared the codes across participants, grouped them into themes aligned with our RQs, and discussed disagreements until reaching consensus.

\subsection{Findings}
\label{sec:qual_findings}
We report four core findings. The first two explain why maintaining the in-house harness became harder to justify and how adopting a general-purpose harness changes engineering responsibilities (RQ3). The third examines the resulting trade-offs in engineering effort, execution cost, runtime control, and latency (RQ3, RQ4). The fourth identifies what must be preserved for autonomous testing to remain trustworthy (RQ4). Each finding ends with an implication for teams in comparable industrial settings; as these findings are grounded in one specific industrial case, we present them as lessons rather than as general best practices.

\emph{F1: Debt emerged from a shifting ecosystem, not a decaying system.} The ecosystem around Setup~A changed in three ways. First, model changes could require renewed validation and adjustments to prompts and agent behavior, although the team sometimes deferred upgrades to limit this burden (Section~\ref{sec:arch_ownership}). D1 described the team's development capacity as insufficient to keep pace with advances in LLMs and agent harnesses. Testers noticed the side effects: after one update, the agent began improperly judging media playback states based on the view hierarchy, failing previously passing tasks (T5). Second, requirements grew beyond the original design, such as operating several devices in one test (D3) and connecting adjacent validation tasks, such as audio, signals, and other displays (PO). Some of these requests emerged because an AI-based platform already existed, not because they served its original purpose of GUI testing (PO). Third, and most critically, general-purpose harnesses matured. D3 explained that Setup~A was built because no capable general-purpose harness existed; once such harnesses appeared, the team was maintaining generic agent infrastructure that vendors now maintain for everyone. D2 and the PO drew the same conclusion: if a vendor-maintained harness performs comparably, maintaining an equivalent in-house is hard to justify. The interviews did not identify deterioration of the original architecture as the primary motivation for considering replacement. Indeed, Setup~A matches Setup~B in task success on the IVI benchmark (Section~\ref{sec:rq1}). Instead, participants emphasized the growing maintenance burden and the emergence of externally maintained alternatives. Thus, the debt lies not in the system's effectiveness, but in the increasing opportunity cost of maintaining its in-house orchestration. We characterize this as \emph{ecosystem-induced technical debt}: a previously justified architecture becomes increasingly costly to maintain relative to external alternatives as the surrounding technological ecosystem matures. 

\emph{Implication.} Revisit the build-versus-adopt decision whenever a new model generation or a mature general-purpose harness becomes available, weighing the cost of keeping the in-house harness current against adopting the external one.

\emph{F2: External harnesses shift maintenance from orchestration to integration.} In Setup~A, extending the harness is engineering work. D3 explained that adding even a second way to tap an element, by coordinates or by text, expands the agent's action choices and may affect existing behavior, so every extension requires re-evaluation; larger requirements, such as operating several devices, conflict with assumptions built into the loop. The developers who had worked with both setups (D2--D4) found Setup~B easier to build, maintain, and extend, as new capabilities are added as tools or instructions. The remaining effort concentrates on the interface: D2 identified the design of clear tool descriptions and parameters as an important integration challenge, since the generic agent must select and invoke the appropriate tools. The PO placed this change in a longer shift of maintenance ownership. With scripted tests, testers repaired their own scripts. With Setup~A, testers report agent failures, the PO prioritizes them, and the AI team fixes the harness. As a result, Setup~A exchanged distributed script maintenance for centralized agent maintenance, which competes with feature requests for the team's capacity. With Setup~B, the generic part of this maintenance moves to the vendor, while the team retains testing-specific instructions, tools, and integration logic.

\emph{Implication.} Measure maintenance effort in both the QA team and the platform team, not only the script maintenance saved by testers. When adopting a vendor-maintained harness, budget for the integration work that remains: the instruction template, the tool descriptions, and the tools.

\emph{F3: Outsourcing orchestration reduces engineering effort but introduces runtime trade-offs.} While Setup~B reduces the internal orchestration effort discussed in F2, it introduces trade-offs in runtime control and execution cost. In Setup~A, the team decides what enters the primary agent's context, what is removed, and how much history is retained (Section~\ref{sec:arch_inhouse}). In Setup~B, several of these decisions are delegated to the external harness. D3 identified context management as the main capability lost in the transition, arguing that the external harness's context-management behavior was not sufficiently tailored to testing. D2 highlighted context compaction during long tasks, whose timing and effects the team cannot directly control; D4 noted that the external harness has greater freedom over its internal execution.

Despite the reduced runtime control, Setup~B achieved comparable or higher aggregate task success in our benchmark (Section~\ref{sec:rq1}). However, it incurred higher inference costs (Section~\ref{sec:rq3_quant}), which developers partly attributed to the larger execution histories repeatedly included in model calls. Consistent with these measurements, developers rated cost as the property with the largest disadvantage for Setup~B. Thus, reducing internal engineering effort may increase the inference expenditure incurred during repeated regression testing.

Execution latency presents a different trade-off. With the same model, Setup~B was at most 11\% faster on the IVI benchmark (Section~\ref{sec:rq3_quant}), yet execution time remained a major concern among testers (T2). T1 estimated that approximately 60 tests required five hours, consistent in magnitude with the benchmark's mean latency of 297~s per task (Table~\ref{tab:efficiency}), although the workloads are not directly equivalent. The PO further emphasized that long-running suites occupy scarce test rigs and delay QA feedback. Nevertheless, testers valued the ability to run tests unattended overnight (T3). Consequently, reducing harness maintenance does not by itself eliminate the operational bottleneck of long-running agentic tests.

\emph{Implication.} Evaluate the total cost of ownership by comparing internal engineering effort with per-run inference cost and expected test frequency. Benchmark representative workloads and track suite duration per rig as a capacity metric; changing the harness alone should not be assumed to resolve execution-time constraints.

\vspace{2mm}
\noindent\fbox{%
    \parbox{\dimexpr\linewidth-2\fboxsep-2\fboxrule}{%
        \textbf{RQ3 findings (maintenance, qualitative):}
        In our case, maintaining the in-house harness became harder to justify as external harnesses matured, creating what we characterize as ecosystem-induced technical debt. Adopting a general-purpose harness reduced internal orchestration responsibilities but introduced trade-offs in runtime control and execution cost, although total cost decreased.
    }%
}

\vspace{2mm}

\emph{F4: Test fidelity requires bounded autonomy and execution visibility.} Testers ranked visibility as Setup~A's strongest property. Autonomous execution has not eliminated human verification: testers described inspecting screenshots, logs, and execution traces to validate failures, sometimes reproducing the workflow manually (T1, T3, T4). Testers also reported behavioral changes following agent updates: after a model update, T5 observed that when an application crash prevented the intended workflow, the agent continued attempting the task instead of recognizing and reporting the blocking failure. Such behavior risks obscuring genuine application defects. Together, the accounts of T1, T3, T4, and T5 suggest that testers value autonomous navigation and adaptation but expect the agent to remain faithful to the specified test scenario.

\emph{Implication.} The goal of a testing agent is to verify the specified path, not only to reach the final state. Regardless of the harness, teams should instruct or constrain the agent to stop and report when the app under test fails instead of working around the failure.

\vspace{2mm}
\noindent\fbox{%
    \parbox{\dimexpr\linewidth-2\fboxsep-2\fboxrule}{%
        \textbf{RQ4 findings:}
        Developers prioritize maintainability, while testers need visible execution and faithful adherence to test scenarios. The autonomy that helps an agent complete open-ended tasks can undermine regression testing when recovery bypasses a defect. Therefore, a replacement must preserve testers' ability to verify both the outcome and the path taken to reach it.
    }%
}

\section{Discussion and Industrial Lessons Learned}
\label{sec:discussion}

Our results suggest that choosing between a custom-built and a general-purpose harness is less a question of which one performs better than of which costs, risks, and responsibilities an organization prefers to carry. We discuss four themes that cut across our quantitative and qualitative results and conclude each with a lesson for practitioners.

\emph{Rethinking technical debt in the LLM era.} Technical debt is usually attributed to expedient decisions within a system~\cite{li2015systematic}, and machine learning systems are known to accumulate debt in the infrastructure around the model~\cite{sculley2015hidden}. Our case points to a further mechanism. Setup~A did not deteriorate: its code did not degrade, and it still matches Setup~B on the IVI benchmark (Section~\ref{sec:rq1}). Instead, its value declined relative to the alternatives as new model generations, new requirements, and mature general-purpose harnesses emerged (F1). Because models evolve much faster than conventional dependencies, such external changes can quickly turn a sound investment into debt.
Adopting a general-purpose harness does not retire this debt but changes its form. The vendor now decides when the loop, the context management, and the compaction policy change, and such updates can alter the agent's behavior without any change on the team's side (F3). The MCP server, the instruction template, and the tool descriptions remain in-house and need upkeep as models and tools evolve (F2). Therefore, the build-versus-adopt decision is not made once: it shifts from maintaining an agent runtime to monitoring an external dependency.
\textbf{Lesson:} Treat an in-house agent harness as a depreciating asset: keep domain knowledge, i.e., tools, instructions, and test intents, separable from the orchestration, so that the orchestration can be replaced once it no longer pays off.

\emph{Trading engineering effort for inference cost.} 
General-purpose harnesses remove what developers described as Setup~A's most demanding maintenance work: retuning the loop, prompts, and context construction for each model generation. New requirements are mostly handled through instructions and tools rather than loop changes (Section~\ref{sec:arch_ownership}, F2). However, inference costs are 1.9--10.7 times as high per task with the same model (2.6--10.7 on the IVI benchmark). Therefore, adoption shifts a largely fixed engineering burden, incurred when models or requirements change, to a variable operational expense paid on every run.
In our case, engineering effort falls from about 0.5 full-time equivalents (FTE) across four developers for Setup~A to 0.1~FTE for Setup~B's tools and instructions. At 60 IVI tasks per working night, Setup~B's additional inference cost over Setup~A's production configuration equals 0.06--0.39~FTE, below the 0.4~FTE saved. Total monthly engineering and inference costs are therefore 3--66\% lower, with OpenCode yielding the largest savings and Claude Code the smallest.
The preferable burden depends on the workload. Execution costs may dominate for large, frequently run stepwise regression suites, where both setups perform comparably. Open-ended tasks or limited developer capacity, as in our case (F1), may justify offloading orchestration. Compact tool outputs can reduce Setup~B's costs because most tokens are re-sent context; OpenCode already costs less than Claude Code at similar success.
Scope creep favors the general-purpose harness. Once an agent works, stakeholders ask it to do more, such as operating several devices or validating audio and signals (F1). A custom-built harness must change its agent loop for such requests, whereas a general-purpose one can often absorb them through an updated instruction template (F2). Still, no harness excels at everything: teams must decide which properties matter most, e.g., success, cost, or control, and accept trade-offs in the others.
\textbf{Lesson:} Decide per test category which burden to carry, engineering effort or inference cost, and scope the agent deliberately, particularly when you are building a custom harness.

\emph{Flexible intents, stable verdicts.} Most IVI intents are written as numbered steps, a style that testers likely carry over from scripted testing: specifying each step ensures that the agent follows the intended path. Other stakeholders prefer goal-level intents and leave the planning to the agent, which is where the general-purpose harnesses performed best (Section~\ref{sec:rq1}). Harnesses should accommodate this full range, including high-level goals with mandatory steps, without imposing one style.
However, higher success on open-ended tasks did not imply greater stability on structured tests. On the IVI benchmark, inconsistent outcomes occurred on 2--3\% of tasks with Setup~A and 5--9\% with Setup~B, although Setup~B was more stable on Google Maps. Regression testing requires consistent verdicts across runs. Testers already verify failures manually (F4), so additional inconsistency increases their workload; even a small stability loss on structured tests may outweigh greater success on open-ended tasks.
\textbf{Lesson:} With natural-language intents, users expect an agent to handle both stepwise procedures and open-ended, high-level goals. Evaluate a candidate harness on both styles and, beyond its success rate, on its stability, i.e., whether repeated runs of the same test yield the same verdict.

\emph{Keeping visibility and control when outsourcing the loop.} Moving to a general-purpose harness outsources the loop but not the responsibility for test results. Testers value seeing what the agent did and need it to stay within the test scenario (F4). The recovery advantage of Setup~B (Section~\ref{sec:rq1}) concerns the agent's own incorrect actions rather than failures of the app under test, and it stays on the intended path: notably, on the AMEX benchmark, recovered runs reached as many milestones as runs without an incorrect call. With a vendor-maintained harness, both must be built deliberately on top of it: run records that the team controls, instructions that require stopping and reporting when the app under test fails, and verdicts that flag workarounds.
A further mechanism is an \emph{experience library}: a curated, team-owned collection of successful and failed trajectories from past tasks, including their intermediate steps~\cite{zhao2024expel,zheng2024synapse}. Supplied to the agent as in-context examples~\cite{brown2020language}, successful trajectories demonstrate the expected path, while annotated failed runs let the agent learn from mistakes in context~\cite{shinn2023reflexion}. Beyond potentially stabilizing behavior and reducing latency and cost on known tasks, such a library makes the expected behavior visible and reviewable for testers. As current models are optimized for instruction following, encoding instructions as examples is also more likely to survive model and harness updates than prompts tuned to a specific model. Such a library pays off mainly in narrow domains with recurring tasks, such as a regression suite for a fixed set of applications, rather than across the diverse applications and intents of our evaluation. We see this as a promising direction for keeping domain knowledge in-house while the orchestration is outsourced.
\textbf{Lesson:} Outsourcing the loop does not outsource trust: give testers what they need to verify the agent, such as team-controlled run records and reviewable examples of expected and failed behavior.

\section{Threats to Validity}
\label{sec:threats}

\emph{Construct validity.} Inference costs reflect vendor prices at the time of our experiments and will change as prices evolve; token counts depend on the tokenizer and are therefore comparable only within a model. In the interviews, the ratings served as prompts for discussion rather than as measurements; we therefore do not report them, and the rated properties differed between developers and testers.

\emph{Internal validity.} 
All participants work with Setup~A, and developers rated Setup~B relative to it by design (anchor-and-compare), anchoring assessments both positively and negatively. Three developers (D2--D4) are co-authors involved in developing both setups and running the quantitative experiments, potentially biasing their accounts; they did not participate in the interview design and analysis.

\emph{External validity.} We selected AMEX applications with dynamic settings resembling our industrial context, but results may differ for other applications.

\section{Conclusion}
\label{sec:conclusion}

Our mixed-methods study combines quantitative benchmarking and qualitative practitioner interviews at Volvo Group to examine the adoption of general-purpose agent harnesses. These harnesses perform better on open-ended tasks and reduce orchestration work. For nightly IVI regression, engineering savings outweigh higher inference costs, lowering total monthly costs by 3--66\%, at the price of less control over what the model receives. 
The custom-built harness remains competitive and more consistent on stepwise IVI tests; it became debt not by deteriorating but because its ecosystem matured, which we term \emph{ecosystem-induced technical debt}.
For workloads like ours, adopting is the better choice; keeping a custom harness pays off mainly where consistency on stepwise tests or control over the model's context matters more than maintenance effort. Teams should keep domain tools and test knowledge separate from the orchestration, so that it can be replaced, and must ensure, whatever the harness, that testers can see what the agent did and that it stays within the test scenario. Future work will explore team-owned libraries of verified execution trajectories to make agent behavior more consistent and reduce token use.

\section*{Acknowledgment}
This work was partially supported by the Wallenberg AI, Autonomous Systems and Software Program (WASP), funded by the Knut and Alice Wallenberg Foundation, and by the Gender Initiative for Excellence (Genie) at Chalmers University of Technology, funded by the Chalmers University Foundation.

\end{document}